\documentclass[aps,pra,twocolumn,amsmath,showpacs,superscriptaddress,tightenlines,pdflatex,longbibliography]{revtex4-1}

\usepackage{amssymb}
\usepackage{amsmath}
\usepackage{dcolumn}
\usepackage{graphicx}
\usepackage{mathrsfs}
\usepackage{appendix}
\usepackage{graphicx}
\usepackage{booktabs}
\usepackage{color}
\usepackage{url}
\usepackage[colorlinks]{hyperref}

\newcommand{\ket}[1]{\mbox{$|#1\rangle$}}
\newcommand{\bra}[1]{\mbox{$\langle#1|$}}

\hypersetup{    plainpages=true,
    breaklinks=true,    hypertexnames=false,    pageanchor=true,
    colorlinks=true,
    linkcolor={blue},
    citecolor={blue},
    urlcolor={blue},
        anchorcolor={black}
}

\begin{document}

\title{Two-color electromagnetically induced transparency via modulated coupling
 between a mechanical resonator and a qubit}

\author{Xin Wang}
\affiliation{Institute of Quantum Optics and Quantum Information,
School of Science, Xi'an Jiaotong University, Xi'an 710049, China}
\affiliation{Theoretical Quantum Physics Laboratory, RIKEN
Cluster for Pioneering Research, Wako-shi, Saitama 351-0198,
Japan}

\author{Adam Miranowicz}
\affiliation{Theoretical Quantum Physics Laboratory, RIKEN
Cluster for Pioneering Research, Wako-shi, Saitama 351-0198,
Japan} \affiliation{Faculty of Physics, Adam Mickiewicz
University, 61-614 Pozna\'n, Poland}

\author{Hong-Rong Li}
\affiliation{Institute of Quantum Optics and Quantum Information,
School of Science, Xi'an Jiaotong University, Xi'an 710049, China}

\author{Fu-Li Li}
\affiliation{Institute of Quantum Optics and Quantum Information,
School of Science, Xi'an Jiaotong University, Xi'an 710049, China}

\author{Franco Nori}
\affiliation{Theoretical Quantum Physics Laboratory, RIKEN
Cluster for Pioneering Research, Wako-shi, Saitama 351-0198,
Japan} \affiliation{Physics Department, The University of
Michigan, Ann Arbor, Michigan 48109-1040, USA}
\date{\today}

\begin{abstract}
We discuss level splitting and sideband transitions induced by a
modulated coupling between a superconducting quantum circuit and a
nanomechanical resonator. First, we show how to achieve an
unconventional time-dependent longitudinal coupling between a flux
(transmon) qubit and the resonator. Considering a sinusoidal
modulation of the coupling strength, we find that a first-order
sideband transition can be split into two. Moreover, under the
driving of a red-detuned field, we discuss the optical response of
the qubit for a resonant probe field. We show that level splitting
induced by modulating this longitudinal coupling can enable
two-color electromagnetically induced transparency (EIT), in
addition to single-color EIT. In contrast to standard predictions
of two-color EIT in atomic systems, we apply here only a single
drive (control) field. The monochromatic modulation of the
coupling strength is equivalent to employing two
eigenfrequency-tunable mechanical resonators. Both drive-probe
detuning for single-color EIT and the distance between transparent
windows for two-color EIT, can be adjusted by tuning the
modulation frequency of the coupling.
\end{abstract}

\pacs{42.50.Ar, 42.50.Pq, 85.25.-j} \maketitle

\affiliation{Institute of Quantum Optics and Quantum Information,
School of Science, Xi'an Jiaotong University, Xi'an 710049, China} %
\affiliation{RIKEN, Wako-shi, Saitama 351-0198, Japan}

\affiliation{RIKEN, Wako-shi, Saitama 351-0198, Japan}
\affiliation{Faculty of Physics, Adam Mickiewicz University,
61-614 Pozna\'n, Poland}

\affiliation{Institute of Quantum Optics and Quantum Information,
School of Science, Xi'an Jiaotong University, Xi'an 710049, China}

\affiliation{RIKEN, Wako-shi, Saitama 351-0198, Japan}
\affiliation{Physics Department, The University of Michigan, Ann
Arbor, Michigan 48109-1040, USA}

\affiliation{Institute of Quantum Optics and Quantum Information,
School of Science, Xi'an Jiaotong University, Xi'an 710049, China} %
\affiliation{RIKEN, Wako-shi, Saitama 351-0198, Japan}

\affiliation{RIKEN, Wako-shi, Saitama 351-0198, Japan}
\affiliation{Faculty of Physics, Adam Mickiewicz University,
61-614 Pozna\'n, Poland}

\affiliation{Institute of Quantum Optics and Quantum Information,
School of Science, Xi'an Jiaotong University, Xi'an 710049, China}

\affiliation{RIKEN, Wako-shi, Saitama 351-0198, Japan}
\affiliation{Physics Department, The University of Michigan, Ann
Arbor, Michigan 48109-1040, USA}

\section{Introduction}

Superconducting quantum circuits
(SQCs)~\cite{Makhlin01,devoret2004,You2005, rClarke08,You2011,
rBuluta11,Xiang13,Schoelkopf2008} are ideal artificial platforms
for studying microwave photonics~\cite{Gu2017,Kockum2018}. Many
quantum-optical effects, such as quantum Rabi
oscillations~\cite{Johansson06,Garziano15}, electromagnetically
induced transparency (EIT)~\cite{Murali04,Dutton06,Ian10,Chang2011,Jing2015,Gu16},
Autler-Townes splitting~\cite{Mika09,Li2012,Novikov13,Suri2013},
and photon blockade~\cite{Hoffman11,Lang11}, have been
successfully demonstrated with SQCs.

In contrast to natural atoms and optical cavities, the basic
elements (i.e., multi-level superconducting systems and
resonators) in SQCs can freely be designed and controlled for
various purposes in microwave photonics~\cite{Gu2017} and
quantum-information technologies~\cite{Wendin2017}.

Most commonly, the coupling between a superconducting qubit and a
single-mode microwave resonator field is
transverse~\cite{You2003,Gu2017}, where the dipole moment of
the qubit interacts with the electric (or magnetic) field of the
resonator mode, and therefore being an exact analog of the
standard quantum Rabi model in cavity quantum electrodynamics
(QED) experiments~\cite{Scully1997}. The quantum dynamics of such
systems has been extensively studied for decades due to its
potential applications in, e.g., quantum information processing
and quantum
optics~\cite{Liu06,Liu2004,Fink2008,Hofheinz2008,WangH2008}.

In recent years, some theoretical and experimental research
studies have been devoted to SQCs with another type of interaction
form, i.e., the so-called \emph{longitudinal}
coupling~\cite{Kerman2013,Liu2014a,Zhao15,Richer16,Richer17}. In a
circuit-QED system with \emph{longitudinal} interaction, the
qubit-transition frequency is modulated by a quantized field, and
the Pauli $\sigma_{z}$ qubit operator couples with a quadrature
field operator~\cite{Richer17}. Compared with the transverse
coupling~\cite{Stassi2018}, the longitudinal coupling has its
inherent advantages since the interaction term commutes with the
qubit operator $\sigma_{z}$. For example, there is no Purcell
decay and residual interactions between a qubit and its resonator.
By using SQCs with longitudinal-coupling, one can realize various
quantum-control tasks, such as error-correction
codes~\cite{Fowler12,Stassi2017} or multiexcitation
generation~\cite{Garziano15,Wangx17,Stassi2017,Kockum2017} among
many other applications~\cite{Gu2017}.

Recently, several studies have been focused on systems with
parametrically-modulated \emph{longitudinal} coupling, where the
interaction strength was rather not constant but modulated in time
at certain frequencies. This modulated interaction can be viewed
as a qubit-state-dependent drive on a resonator. If the modulation
rate is equal to the resonator frequency, qubit states can be
readout rapidly via quantum nondemolition (QND)
measurements~\cite{Didier15}. Moreover, it is possible to obtain a
high-fidelity controlled-phase gate, if modulating the
longitudinal coupling between two remote qubits and a common
resonator~\cite{Royer2017}. All these studies indicate that the
\emph{modulated longitudinal} coupling has its own advantages over
the transverse and \emph{constant} longitudinal interactions, and
provides another way to achieve better quantum control and
engineering. However, as discussed in
Refs.~\cite{Didier15,Liao16,Cirio17}, it is not easy to obtain
such modulated couplings in either natural or artificial systems.

In this paper, we  describe possible sideband transitions and the
optical response in a system with a parametrically-modulated
longitudinal coupling which, to our knowledge, has not been
discussed in previous studies. We start our discussions by
proposing two possible circuit layouts, where superconducting
qubits are longitudinally coupled to nanomechanical resonators
(NAMRs) via an external flux~\cite{Shevchuk17}. The coupling can
be conveniently modulated in time by changing external magnetic
fields. Considering a transverse driving field of a qubit, we find
that a sideband transition can be split into several asymmetric
parts if the modulation is sinusoidal.

Assuming that a resonant probe field is also applied to a qubit,
we demonstrate that both single- and two-color (bichromatic)
EIT~\cite{Wang03,WangH14,Yan13} can be observed. For single-color
EIT, parametric modulation is equal to a flexible NAMR with a
tunable eigenfrequency and, therefore, the drive-probe detuning of
the EIT dip can be conveniently tuned. For two-color EIT, two
transparent windows result from the parametric modulation of the
longitudinal coupling, and the distance between two transparent windows can be adjusted by changing the modulation frequency,
rather than sweeping two control frequencies in a conventional
bichromatic EIT system~\cite{Wang06,Moiseev06,Yan13}.

There are various potential applications of EIT based on
circuit-QED systems, such as optical
switching~\cite{Bajcsy09,Xia13,Gu2017}, controlling slow light for
information storage, demonstrating single-photon router
devices~\cite{Hoi11,Leung12}, and controlling photon transmission
through a circuit-QED system~\cite{He2007,Liu2014b}. Two-color EIT can be
employed for entangling photons via cross-phase modulation and
slowing photons at different
frequencies~\cite{Lukin2000,Wang06,Lisj08}. Our results can be
helpful to study the dynamics for systems with time-dependent
longitudinal coupling, and applications based on EIT in microwave
photonics~\cite{Gu2017}.

The outline of the paper is as follows. In Sec.~II, we describe a
possible approach to mediate a flux (transmon) qubit with a NAMR
via a modulated longitudinal coupling.  In Sec.~III, we
derive an analytical Hamiltonian describing sideband-transition
splitting. In Sec.~IV, we discuss single- and two-color EIT, and
show how to tune these two effects by changing drive-field
parameters. Our final discussions and conclusions are
presented in Sec.~V.

\section{Model}

A possible circuit-QED implementation of the time-dependent
longitudinal interaction has been discussed in
Ref.~\cite{Didier15}. Specifically, a transmon qubit was assumed
to interact with a $\lambda/4$ transmission-line resonator via a
Josephson junction inserted at the end of a central conductor. By
modulating an external flux through the superconducting quantum
interference device (SQUID) loop of the transmon qubit, a desired
sinusoidally-modulated coupling can be obtained. However, the
qubit transition frequency is also perturbed by a time-dependent
control flux (with a frequency range about tens of MHz), which should
be avoided in certain cases.

Here we demonstrate another possible hybrid circuit layout to
realize such modulated interactions between high-frequency NAMR
and superconducting qubits.

\subsection{Flux-mediated coupling between SQUID and NAMR}
\begin{figure}[t]
\centering \includegraphics[width=6.0cm]{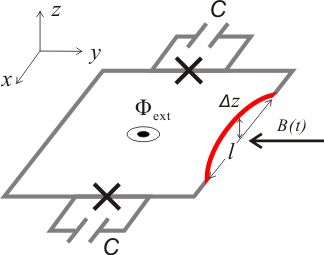} \caption{Two
Josephson junctions with an identical capacitance $C$ form a SQUID
loop in the $x$-$y$ plane. A NAMR (represented by the red bow-shaped curve),
with length $l$ and mass $m$, is embedded in the loop and vibrates
in the $z$-direction with amplitude $\Delta z$. The static
magnetic field perpendicular to the $x$-$y$ plane produces a
static flux $\Phi_{\rm{ext}}$ for the SQUID. Additionally, there
is a time-dependent magnetic field $B(t)$ [parallel to the $x$-$y$
plane] applied to the loop area produced by the NAMR. The total
flux through the SQUID is modulated by the vibrational motions of
NAMR. } \label{fig1}
\end{figure}
We start our discussions by describing the interaction between a
SQUID and a NAMR. As shown in Fig.~\ref{fig1}, two Josephson
junctions, each shunted by a capacitance $C$, are connected by a
loop in the $x$-$y$ plane. The total charging energy for two
junctions is $E_{c}=e^{2}/C$. The NAMR with length $l$ is
coated with a superconductor and fabricated into the
loop~\cite{Etaki2008,Poot2010,Etaki2013}. Alternatively,
carbon nanotubes could be employed (acting as superconducting
junctions \emph{in situ}) to produce mechanical
vibrations~\cite{Schneider12,Shevchuk17}. The NAMR (with mass $m$
and frequency $\omega_{m}$) vibrates along the $z$-direction, and
a time-dependent magnetic field $B(t)$ is perpendicularly applied
to the arm of the NAMR in the
$y$-direction~\cite{Etaki2008,Poot2010}.

Denoting the gauge-invariant phase difference by $\phi_{i}$ and
the Josephson energy by $E_{Ji}$ for the $i$th junction, we can
write the total Josephson energy for the SQUID as ($\hbar =1$):
\begin{eqnarray}
E_{J}&=&E_{J1}\cos\phi_{1}+E_{J2}\cos\phi_{2}   \notag\\
&=&E_{J\Sigma}\big( \cos\phi_{+}\cos\phi_{-}-\ d_0
\sin\phi_{+}\sin\phi_{-}\big), \label{eqjj}
\end{eqnarray}
where $E_{J\Sigma}=E_{J1}+E_{J2}$ is the total Josephson energy,
$\ d_0=(E_{J1}-E_{J2})/E_{J\Sigma}$ is the junction asymmetry, and
$\phi_{+}=(\phi_{1}+\phi_{2})/2$ represents the overall phase of
the SQUID~\citep{Schneider12}. Note that
$\phi_{-}=(\phi_{1}-\phi_{2})/2=\pi\Phi_{x}/\Phi_{0}$ is bound by
the fluxoid quantization relation, where $\Phi_{x}$ is the flux
through the SQUID ring and $\Phi_{0}$ is the flux quantum. Here we
have neglected the geometric and kinetic inductances of the
loop~\cite{Schwarz15doc}. As shown in Fig.~\ref{fig1}, there are
two components for the flux $\Phi_{x}$~\cite{Xue2007b,Shevchuk17}:
The static part $\Phi_{\rm{ext}}$, which is induced by a
homogeneous magnetic field in the $z$-direction, and a
time-dependent part resulting from the NAMR vibrating around its
equilibrium position and the $y$-direction magnetic field $B(t)$.
Therefore $\phi_{-}$ is expressed as
\begin{equation}
\phi_{-}=\frac{\pi\Phi_{x}}{\Phi_{0}}=\frac{\pi}{\Phi_{0}}[\Phi_{\rm{ext}}+B(t)\xi
l\Delta z], \label{eqflux}
\end{equation}
where $\xi$ is the average geometric constant~\cite{Xue2007a}, and
$\Delta z$ is the displacement of the NAMR away from its
equilibrium position at $z=0$.

Note that $\Phi_{\rm{ext}}$ and $B(t)$ are induced by two magnetic
fields with perpendicular directions. The first and second terms
in Eq.~(\ref{eqflux}) can, in principle, be changed independently.
An imperfect perpendicular relation between these two magnetic
fields might cause the net control flux, which should be minimized
in experiments. To obtain the coupling relation between the
vibration mode and the SQUID, we rewrite Eq.~(\ref{eqjj}) as
\begin{equation}
E_{J}=E_{J\Sigma}^{\prime}\cos(\phi_{+}+\phi_{0}), \label{EJX}
\end{equation}
where $E_{J\Sigma}^{\prime}=E_{J\Sigma}\sqrt{\cos^{2}{\phi_{-}}+\
d_0^{2}\sin^{2}{\phi_{-}}}$ is the effective Josephson energy, and
$\phi_{0}=\arctan(d_0 \tan{\phi_{-}})$ is the shifted phase. We
assume that the junction asymmetry $\ d_0\ll1$, and $\phi_{-}$ is
far away from $\pi/2$. As a result, $\phi_{0}$ is only a small
constant phase factor, which has a small effect on the kinetic
energy~\citep{Shevchuk17}. The vibrational motion of the NAMR
induces a flux perturbation on $E_{J\Sigma}^{\prime}$. By
expanding the displacement-dependent $E_{J\Sigma}^{\prime}$ to
first order in $\Delta z$, we obtain
\begin{eqnarray}
\frac{\partial E_{J\Sigma}^{\prime}}{\partial \Delta
z}&=&\frac{\partial E_{J\Sigma}^{\prime}}{\partial \phi_{-}}
\frac{\partial\phi_{-}}{\partial \Delta z} \notag \\
&=&-\frac{\pi E_{J\Sigma}\sin(2\phi_{-})(1-\ d_0^{2})B(t)\xi
l}{2\Phi_{0}\sqrt{(1-\ d_0^{2})\cos^{2}\phi_{-}+\ d_0^{2}}}.
\label{eqfluxbias}
\end{eqnarray}
Considering that two junctions are symmetric with
$d_0=0$, we reduce Eq.~(\ref{eqfluxbias}) to a simpler
form
\begin{equation}
\frac{\partial E_{J\Sigma}^{\prime}}{\partial \Delta z}\Big|_{\
d_0=0} =-\frac{\pi E_{J\Sigma}\sin(\phi_{-})B(t)\xi l}{\Phi_{0}},
\label{eqsimple}
\end{equation}
from which we find that the magnetic field $B(t)$, together with
the mechanical oscillations, induce a time-dependent
modulation of the effective Josephson energy
$E_{J\Sigma}^{\prime}$ of the SQUID.

\subsection{Longitudinal interaction between superconducting qubits and NAMRs}

\begin{figure}[t]
\centering \includegraphics[width=8.0cm]{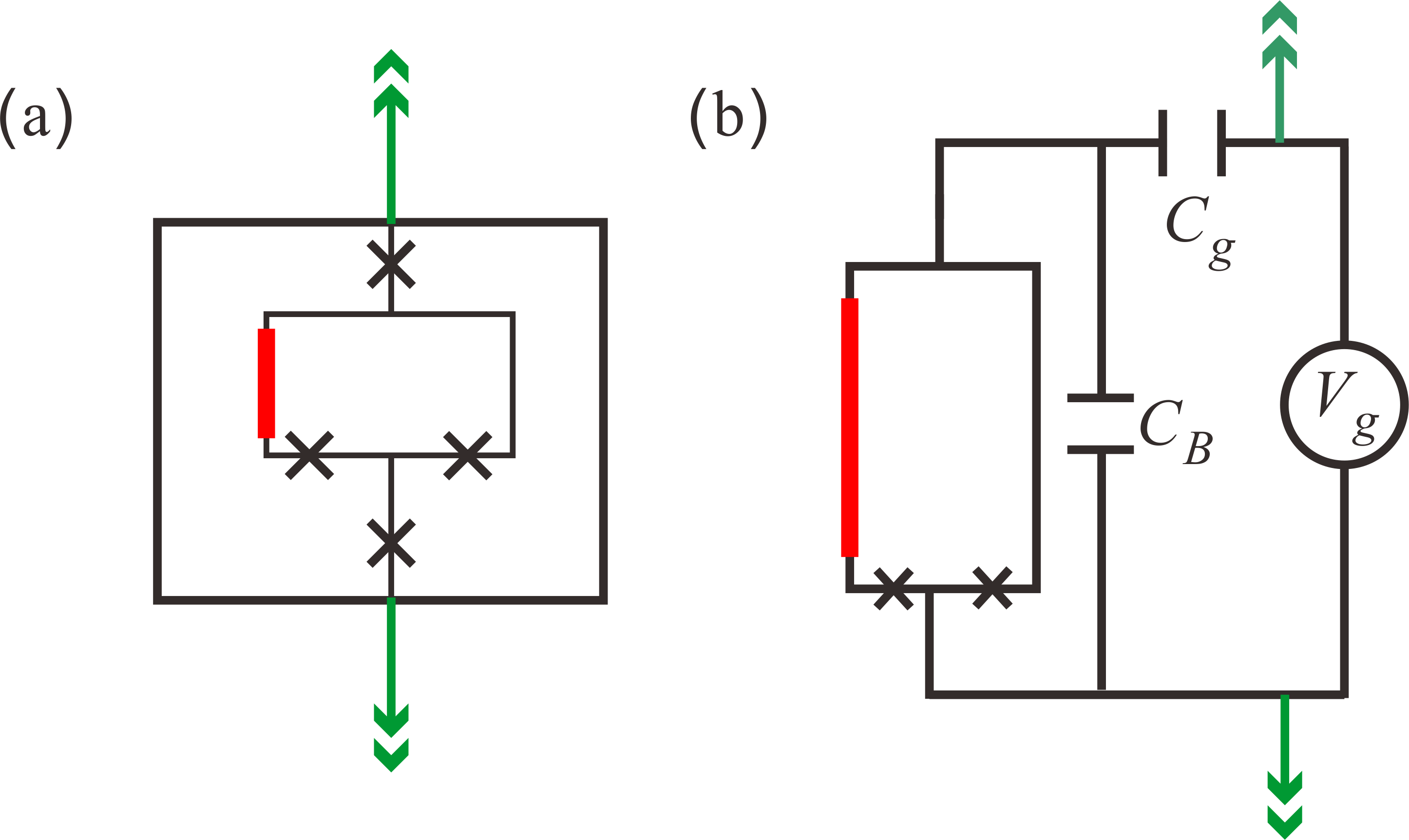} \caption {(a)
Schematic diagram for coupling a NAMR with a flux qubit. The small
$\alpha$-junction for the flux qubit is replaced by the SQUID in
Fig.~\ref{fig1}. The NAMR (the red thick line) vibrates in the direction
perpendicular to this plane. (b) Schematic diagram of a NAMR
coupled to a transmon qubit. Similar to the flux qubit case (a),
the single junction is replaced by the hybrid mechanical-SQUID
system. The time-dependent flux changes the effective Josephson
energy of the transmon qubit. In panels (a) and (b), the coherent
driving fields are applied through the 1D transmission line (green arrows).} \label{fig2}
\end{figure}

In typical circuit layouts of a superconducting qubit, we often
replace a nonlinear Josephson junction with a SQUID for tuning the
Josephson energy, which enables more flexibility and tunability
for controlling qubits. For example, as shown in
Fig.~\ref{fig2}(a), we consider a gradiometric gap-tunable flux
qubit, where the SQUID and two identical junctions with their
Josephson energy $E_{J0}$, form two symmetric gradiometric
loops~\cite{Paauw09,Paauwthesis,Fedorov10,Stern14}. Compared with
a three-junction flux qubit~\cite{Mooij99,Orlando99}, the small
$\alpha$-junction is replaced by a mechanical-SQUID system, where
$\alpha=E_{J\Sigma}^{\prime}/E_{J0}$ is operated in the regime
$0.5<\alpha<1$~\cite{Paauwthesis,Schwarz15doc,Orlando99}. The NAMR
is shown by the red thick line. As demonstrated in
Refs.~\cite{Paauw09,Fedorov10}, around the optimal point, the
Hamiltonian for the flux qubit can be approximately expressed as
\begin{equation}
H_{q}=\frac{1}{2}\Delta_{t} \sigma _{z}+\frac{1}{2}\epsilon \sigma
_{x}, \label{qubit}
\end{equation}
where $\sigma_{z}=\ket{e}\bra{e}-\ket{g}\bra{g}$ and
$\sigma_{x}=\ket{e}\bra{g}+\ket{g}\bra{e}$ are Pauli operators,
given in terms of the ground $\ket{g}$ and excited $\ket{e}$
states of the qubit. Moreover, the energy bias
$\epsilon=2I_{p}\delta\Phi_{q}$ is controlled via the imbalance
flux $\delta\Phi_{q}$ through the two gradiometric loops, and
$I_{p}$ is the persistent-current amplitude. In experiments,
$\delta\Phi_{q}$ can be induced via a propagating microwave field
in a 1D transmission line [green line in Fig.~\ref{fig2}(a)].

The energy gap $\Delta_{t}$ depends on the parameter $\alpha$.
Specifically, the following approximate analytical expression can
be obtained by tight-binding calculations of the
eigenstates~\cite{Schwarz15doc,Orlando99}:
\begin{eqnarray}
\Delta_{t}&=&\sqrt{\frac{4E_{J0}E_{C}(2\alpha^{2}-1)}{\alpha}} \notag \\
&&\times\exp\left[-g(\alpha)\sqrt{4\alpha(1+2\alpha)\frac{E_{J}}{E_{C}}}
\right], \label{eqDa}
\end{eqnarray}
where $g(\alpha)=\sqrt{1-1/(4\alpha)^{2}}-\arccos[1/(2\alpha)]/(2\alpha)$.
Note that $\alpha$ linearly depends on the effective Josephson energy
$E_{J\Sigma}^{\prime}$ of the SQUID. Therefore, we can express the
sensitivity $R_{f}$ of the gap $\Delta_{t}$ to the flux control
$\phi_{-}$ through the SQUID loop as
follows~\cite{Paauwthesis,Schwarz15doc}
\begin{equation}
R_{f}=\frac{\partial \Delta_{t}}{\partial \phi_{-}}=\frac{\partial
\Delta_{t}}{\partial \alpha} \frac{\partial \alpha}{\partial
E_{J\Sigma}^{\prime}} \frac{\partial
E_{J\Sigma}^{\prime}}{\partial \phi_{-}},
\end{equation}
where $\partial \Delta_{t}/\partial \alpha$ is obtained from
Eq.~(\ref{eqDa}), and $\partial E_{J\Sigma}^{\prime}/\partial
\phi_{-}$ is given in Eq.~(\ref{eqfluxbias}). Indeed, the flux
sensitivity $R_{f}$ could be directly obtained in experiments by
detecting a qubit spectrum via sweeping the flux control
$\phi_{-}$. As shown in Refs.~\cite{Paauwthesis,Paauw09}, $R_{f}$
is about $0.07\sim0.7~\rm{GHz}/{m\Phi_{0}}$. The flux perturbation
results from the vibrations of the NAMR and the time-dependent
magnetic field $B(t)$~\cite{Wang17b}, and therefore, $\Delta_{t}$
can be approximately rewritten as
\begin{eqnarray}
\Delta_{t}&=&\Delta_{t}(z=0)+R_{f} \frac{\partial \phi_{-}}{\partial \Delta z}\Delta z \notag \\
&=&\Delta_{t}(z=0)+R_{f} \frac{\pi B(t)\xi l}{\Phi_{0}} \Delta z.
\label{dtx}
\end{eqnarray}
where $\Delta_{t}(z=0)$ is the energy gap of the qubit when the NAMR is at its equilibrium position $z=0$.

Following Eqs.~(\ref{qubit})-(\ref{dtx}) and considering the NAMR
free energy, the Hamiltonian of this hybrid system becomes
\begin{equation}
H_{c}=\frac{1}{2}\Delta_{t}(z=0) \sigma _{z}+ \frac{1}{2}\epsilon
\sigma _{x} +\omega _{m}b^{\dag }b+g(t)\sigma _{z}(b^{\dag }+b),
\label{eqcoupling1}
\end{equation}
where $b$ ($b^{\dag}$) is the phonon annihilation (creation)
operator of the NAMR. By expanding $B(t)$ with its Fourier
transform of frequency $\omega^{\prime}$, the
longitudinal-interaction strength can be written as
\begin{equation}
g(t)=R_{f} \frac{\pi\xi l}{\Phi_{0}} x_{0}\sum_{\omega^{\prime}}
\big[B(\omega^{\prime}) e^{-i\omega^{\prime}
t}+B^{\star}(\omega^{\prime})e^{i \omega^{\prime}t}\big],
\label{eqgt}
\end{equation}
where $x_{0}=\sqrt{1/(2m\omega_{m})}$ is the zero-point
fluctuation of the NAMR, and $|B(\omega^{\prime})|$ is the
magnetic-field amplitude of the frequency $\omega^{\prime}$
component. However, the magnetic noise through the SQUID loop also
causes decoherence of the qubit via the flux sensitivity $R_{f}$
\emph{in situ}~\cite{Paauwthesis,Schwarz15doc,Paauw09}, the
relaxation times (both $T_{1}$ and $T_{2}$) decrease with
increasing $R_{f}$. To suppress these decoherence processes, we
should control $R_{f}$ below a certain level.

Besides employing a flux qubit, it is possible to induce the
time-dependent interaction between a transmon
qubit~\cite{Koch07,You2007,Schreier08,Mallet2009} and a NAMR. As
depicted in Fig.~\ref{fig2}(b), the single Josephson junction for
the transmon qubit is replaced by a SQUID embedded by a NAMR. The
charging energy $E_{C}$ is reduced by adding a large shunt
capacitance $C_{B}$~\cite{Koch07,You2007,Schreier08}, and the
transmon qubit is insensitive to charge noise under the condition
$E_{J\Sigma}^{\prime}\gg E_{C}$. For simplicity, we assume that
the driving field propagating along the 1D transmission line
(plotted as green arrows in Fig.~\ref{fig2}) and the bias voltage
$V_{g}$ are applied to the gate capacitance $C_{g}$. Given that
the transmon qubit, shown in Fig.~\ref{fig2}(b), can be
approximately viewed as a Duffing oscillator, the transition
frequency between the two lowest eigenstates
is~\cite{Schneider12,Koch07}:
\begin{equation}
E_{01}=\sqrt{8E_{C}E_{J\Sigma}^{\prime}}-E_{C}.
\label{eqcoupling2}
\end{equation}
For simplicity, we consider the symmetric junction at $d_{0}=0$,
then the flux sensitivity on $E_{01}$ of the transmon qubit is
\begin{eqnarray}
R_{t}&=&\frac{\partial E_{01}}{\partial \phi_{-}}=\frac{\partial
E_{01}}{\partial E_{J\Sigma}^{\prime}}
\frac{\partial E_{J\Sigma}^{\prime}}{\partial \phi_{-}}  \notag \\
&=&\sqrt{2E_{C}E_{J\Sigma}\sin{\phi_{-}}\tan{\phi_{-}}}.
\end{eqnarray}
By assuming that $E_{J\Sigma}=70~\rm{GHz}$, $E_{C}=2~\rm{GHz}$,
and $\phi_{-}=\pi/3$, we obtain $R_{t}\simeq0.064\,
\rm{GHz}/{m\Phi_{0}}$. Similar to the above discussions about our
derivation of Eq.~(\ref{eqcoupling2}), we can also obtain the
modulated coupling between the transmon qubit and the NAMR.

Here we discuss the coupling strength under current experimental
conditions, and choose a carbon-nanotube resonator as the NAMR,
e.g., with effective mass $m=4\times10^{-21}~\text{kg}$ and
fundamental frequency
$\omega_{m}/(2\pi)=100~\rm{MHz}$~\cite{Yao00,Huttel09,Laird11,Benyamini14}.
The NAMR length can be of $3~\mu\text{m}$ with geometric constant
$\xi=0.9$~\cite{Xue2007a,Etaki2008,Shevchuk17}. Employing the flux
sensitivity $R_{f,t}=0.25~\rm{GHz}/{m\Phi_{0}}$ and the magnetic
field amplitude $|B(\omega^{\prime})|\simeq 800~\mu\text{T}$, the
coupling strength for the frequency component $\omega^{\prime}$ is
about $g(\omega^{\prime})/(2\pi)\simeq 8~\rm{MHz}$.

There are many potential applications for these time-dependent
longitudinally-coupled systems, such as generating
macroscopic nonclassical states and performing quantum
nondemolition measurements of the qubit
states~\cite{Didier15,Royer2017}.

In the following, we discuss the optical response of a microwave
field applied to the qubit, and show how to observe various types
of tunable EIT based on such a hybrid system.

\section{Effective Hamiltonian for tunable sideband transitions}
To observe sideband transitions and EIT, we assume that a strong
drive and a weak probe, with frequencies $\omega _{\text{drv}}$
and $\omega _{\text{pr}}$, respectively, are applied to the
superconducting qubit through the 1D transmission
line~\cite{Astafiev2010,Anisimov2011,Hoi11,Hoi12}.
The coherent drive and probe fields are both approximately
at the qubit transition frequency $\omega_{q}$, which
is usually around several $\text{GHz}$. Therefore the
oscillation frequency of the counter-rotating terms is
about $2\omega_{q}$. In the following discussions,
the Rabi frequencies of the drive and probe fields
are assumed to be within several $\text{MHz}$. Therefore we can
adopt the rotating wave approximation and neglect the counter-rotating terms. Thus,
the driven hybrid system can be written as
\begin{eqnarray}
\bar{H} &=&\frac{1}{2}\omega_{q} \sigma _{z}+\omega _{m}b^{\dag
}b+g(t)\sigma
_{z}(b+b^{\dag })  \notag \\
&&-\left[(\Omega _{\text{drv}}e^{-i\omega _{\text{drv}}t}+\Omega _{\text{%
pr}}e^{-i\omega _{\text{pr}}t})\sigma _{+}+\text{H.c.}\right],
\label{eq1}
\end{eqnarray}%
where $\Omega _{\text{drv}}$ and $\Omega _{\text{pr}}$ are the
Rabi frequencies of the drive and probe fields, respectively,
and $\sigma_{+}=\ket{e}\bra{g}$ ($\sigma_{-}=\ket{g}\bra{e}$)
is the raising (lowering) operator for the qubit.

Applying a
frame rotating at frequency $\omega _{\text{drv}}$, the Hamiltonian in
Eq.~(\ref{eq1}) is transformed to
\begin{eqnarray}
\bar{H} &=&\frac{1}{2}\Delta_{0} \sigma _{z}+\omega _{m}b^{\dag
}b+g(t)\sigma
_{z}(b+b^{\dag })  \notag \\
&&-\Omega _{\text{drv}}(\sigma _{+}+\sigma _{-})-\Omega
_{\text{pr}}(\sigma _{+}e^{-i\delta t}+\sigma _{-}e^{i\delta t}),
\label{eq2}
\end{eqnarray}%
where $\Delta_{0} =\omega_{q} -\omega _{\text{drv}}$ is the
qubit-drive detuning, $\delta =\omega _{\text{pr}}-\omega
_{\text{drv}}$ is the probe-drive detuning, and $\sigma
_{x}=\sigma _{+}+\sigma _{-}$. Assuming that
$\Omega_{\text{drv}}\gg \Omega _{\text{pr}}$, we can neglect the
last term in Eq.~(\ref{eq2}).
By applying the time-dependent polariton transformation $U(t)=\exp
[\sigma_{z}Y(t)]$ to $\bar{H}$ (see, e.g.,~\cite{Royer2017}),
where $Y(t)=\beta ^{\ast}(t)b^{\dag}-\beta(t)b,$ we obtain the
transformed Hamiltonian
\begin{eqnarray}
H&=&U^{\dag}(t)\bar{H}U(t)-iU^{\dag}(t)\frac{\partial U(t)}{\partial t} \notag \\
&=&\frac{1}{2}\Delta_{0} \sigma _{z}+\omega _{m}b^{\dag }b+\sigma_{z}\big[\eta(t)b+\eta^{\ast}(t)b^{\dag }\big] \notag \\
&&-\Omega _{\text{drv}}\left[\sigma
_{+}e^{2Y(t)}+\rm{H.c.}\right], \label{eq4}
\end{eqnarray}
where $\eta(t)=g(t)-\omega_{m}\beta(t)+i\frac {\partial }
{\partial t} \beta(t).$ We can eliminate the longitudinal-coupling
terms in Eq.~(\ref{eq4}) by setting $\eta(t)\equiv0$. Assuming
that $\beta(t)=M(t)+iN(t)$, the following relations should be
satisfied
\begin{align}
0&=-\omega_{m}M(t)+g(t)-\frac {\partial N(t)} {\partial t}, \label{eq5a}\\
0&=-\omega_{m}N(t)+\frac {\partial M(t)} {\partial t}.
\label{eq5b}
\end{align}

To simplify our analysis, we assume that $g(t)$ is sinusoidally
modulated by a monochromatic drive at frequency $\omega_{g}$,
i.e.,
\begin{equation}
  g(t)=g_{0}\cos(\omega_{g}t).
 \label{gt}
\end{equation}
The general solution for the differential Eqs.~(\ref{eq5a}) and
(\ref{eq5b}) has the form
\begin{equation}
\beta(t)=C_{0}e^{-i\omega_{m}t}+\frac{g_{0}}
{\omega_{m}^{2}-\omega_{g}^{2}}\big[\omega_{m}\cos(\omega_{g}t)-i\omega_{g}\sin(\omega_{g}t)\big],
\end{equation}
where $C_{0}$ is an arbitrary complex coefficient. For simplicity,
by setting $C_{0}=0$, $\beta(t)$ is reduced to
\begin{equation}
\beta(t)=\frac{g_{0}}{2}\left(
\frac{e^{i\omega_{g}t}}{\omega_{m}+\omega_{g}}+\frac{e^{-i\omega_{g}t}}{\omega_{m}-\omega_{g}}\right).
\label{eq6}
\end{equation}
Under the condition $\eta(t)=0$, Eq.~(\ref{eq4}) is simplified as
\begin{equation}
H=\frac{1}{2}\Delta_{0} \sigma _{z}+\omega _{m}b^{\dag }b- \Omega
_{\text{drv}}\left[\sigma _{+}e^{2Y(t)}+\rm{H.c.}\right],
\label{eq7}
\end{equation}
The coupling-modulation frequency $\omega_{g}$ and the
coupling constant $g_{0}$ are assumed to be much smaller than the
NAMR frequency, i.e., $\{\omega_{g},g_{0}\}\ll\omega_{m}$.
Therefore, $|\beta(t)|$ is always a small dimensionless
parameter, as it holds
\begin{equation}
|\beta(t)|\leq\frac{g_{0}\sqrt{\omega_{m}^{2}+\omega_{g}^{2}}}{\omega_{m}^{2}-\omega_{g}^{2}}\ll1.
\label{eqbetall}
\end{equation}
Expanding the last term of Eq.~(\ref{eq7}) to the third order in
$\beta(t)$, the Hamiltonian reads
\begin{eqnarray}
H &=&\frac{1}{2}\Delta_{0} \sigma _{z}+\omega _{m}b^{\dag }b-\Omega _{%
\text{drv}}[\sigma _{+}+\sigma _{-}]  \notag \\
&-&2\Omega _{\text{drv}}\Big\{\sigma
_{+}\big[Y(t)+Y^{2}(t)+\frac{2}{3}Y^{3}(t)\big]+\text{H.c.}\Big\}.
\label{eq9}
\end{eqnarray}
We assume a red-sideband drive field with an amplitude, which
is weak compared with the detuning frequency, i.e., $\Omega
_{\text{drv}}\ll\Delta_{0}\simeq\omega_{m}$. Similar to
discussions in Ref.~\cite{Wang16e}, the third term in
Eq.~(\ref{eq9}) leads to a dynamical Stark shift for the qubit. We
can rewrite the Hamiltonian in Eq.~(\ref{eq9}) in the basis of its
eigenstates:
\begin{subequations}
\begin{gather}
|+\rangle=\cos{\frac{\theta}{2}}|e\rangle+\sin{\frac{\theta}{2}}|g\rangle,\\
|-\rangle=\cos{\frac{\theta}{2}}|g\rangle-\sin{\frac{\theta}{2}}|e\rangle,
\end{gather}
\end{subequations}
with $\tan{\theta}=-2\Omega_{\text{drv}}/\Delta_{0}$. Since
$\theta\ll1$, we can neglect the rotating angle of the qubit basis.
The dynamical Stark shift slightly shifts the qubit-transition
frequency. Thus, we should replace the detuning $\Delta_{0}$ in
the qubit free-energy term in Eq.~(\ref{eq9}) with a modified
qubit-drive detuning
\begin{equation}
\widetilde{\Delta}=\sqrt{\Delta_{0}^{2}+4\Omega_{\text{drv}}^{2}}.
\end{equation}
We assume that the drive detuning $\delta_{s}\equiv
\widetilde{\Delta}-\omega_{m}$, is of the same order as
coupling-modulation frequency $\omega_{g}$.
According to Eq.~(\ref{eqbetall}), $|\beta(t)|$ is
a small parameter and it can easily be verified that
$|\beta(t)|^{N}\Omega_{\text{drv}}\ll\omega_{m}\simeq\widetilde{\Delta}$ ($N=1,2,3$).
Under these conditions, we can neglect the
rapidly-oscillating terms in Eq.~(\ref{eq9}).
By applying the
unitary transformation $U_{0}(t)=\exp
[-i(\tilde{\Delta}\sigma_{z}+\omega _{m}b^{\dag }b)t]$ to
Eq.~(\ref{eq9}), we
obtain the effective Hamiltonian
\begin{equation}
H=\xi(t)\sigma _{+}b\exp (i\delta_{s}t)+\rm{H.c.} \label{eq11}
\end{equation}
Note that $\xi(t)$ has been decomposed into the frequency
components $\pm\omega_{g}$ and $\pm3\omega_{g}$ as follows:
\begin{eqnarray}
\xi(t)&=&2\beta(t)\Omega_{\text{drv}}-\frac{4N}{3}\beta(t)^{2}\beta^{\ast}(t)\Omega_{\text{drv}} \notag\\
&=&\sum_{j=\pm}\big(C_{1j}e^{ji\omega_{g}t}+C_{3j}e^{j3i\omega_{g}t}\big)
\label{eq12}
\end{eqnarray}
with the first- and third-order sideband transition rates
given respectively by:
\begin{align}
C_{1\pm}&=\frac{g_{0}\Omega_{\text{drv}}}{\omega_{m}\mp\omega_{g}}
-\frac{4Ng_{0}^{3}\Omega_{\text{drv}}(3\omega_{m}\mp\omega_{g})}{3(\omega_{m}^{2}-\omega_{g}^{2})^{2}}, \label{eq13} \\
C_{3\pm}&=\frac{4Ng_{0}^{3}\Omega_{\text{drv}}}{3(\omega_{m}^{2}-\omega_{g}^{2})(\omega_{m}\mp\omega_{g})},
\label{eq14}
\end{align}
where $N=\langle b^{\dag}b \rangle -1$ with $\langle b^{\dag}b \rangle$ being the average phonon number. The NAMR is assumed to be in the
quantum regime with several phonons, and $N$ is not large. When
deriving Eq.~({\ref{eq11}), the rapidly-oscillating terms
were neglected and only near-resonant ones were kept.
Equation~(\ref{eq9}) was expanded to third order in
$Y(t)$ only, as observable effects of higher-order terms can be
ignored.} If sweeping the sideband drive frequency around the
regime $\tilde{\Delta}\simeq \omega _{m}$, four apparent resonant
positions can be observed at $\delta_{s}=\pm\omega_{g},
\pm3\omega_{g}$. It is easy to find first-order transition
rates
\begin{equation}
C_{1\pm}\simeq\frac{g_{0}\Omega_{\text{drv}}}{\omega_{m}\mp\omega_{g}}
\gg C_{3\pm} \label{eqC1pm}
\end{equation}
by assuming that $|\beta(t)|\ll1$. In some of the following
discussions, we neglect the third-order transition rates
$C_{3\pm}$. To observe the optical response of the probe field, the
probe term should be added (in the rotating frame), then the total
Hamiltonian becomes
\begin{equation}
H=\sum_{j=\pm}\big[C_{1j}\sigma _{+}b e^{
i(j\omega_{g}+\delta_{s})t} -\Omega_{\text{pr}}\sigma _{+}
e^{i(\tilde{\Delta}-\delta) t}\big]+\rm{H.c.} \label{eq15}
\end{equation}

\begin{figure}[t]
\centering \includegraphics[width=8.5cm]{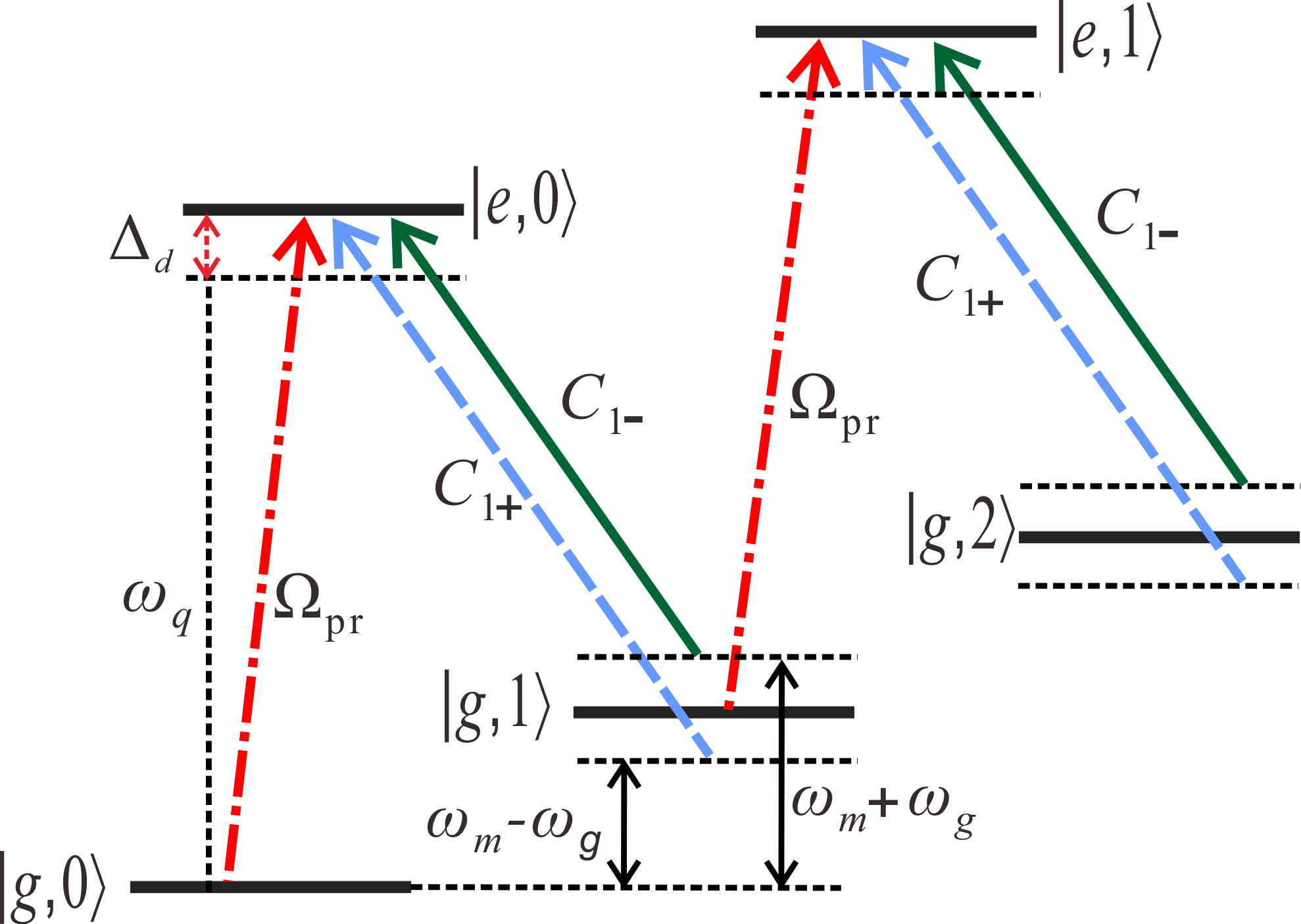}
\caption{Lowest-energy levels of the qubit-NAMR system and
possible state transitions induced by a sideband drive (of
frequency $\Omega_{\rm{drv}}$, which is not shown here) and a
resonant probe (of frequency $\Omega_{\rm{pr}}$, represented by
the red dashed-dotted arrow). The dynamical Stark shift $\Delta_{\rm d}$ results
from the detuning of the drive field. Due to a sinusoidal
modulation of the longitudinal coupling at frequency $\omega_{g}$,
the first-order sideband transition, induced by the drive field
(from $|e,n\rangle$ to $|g,n+1\rangle$), is split into two
transitions with rates $C_{1+}$ (blue dashed arrows) and $C_{1-}$ (green
solid arrows). Note that $\omega_{q}$ ($\omega_{m}$) is the qubit (NAMR)
frequency, and $\omega_{g}$ is the coupling-modulation frequency.}
\label{fig3}
\end{figure}

The energy-level diagram is depicted in Fig.~\ref{fig3}. Due to
the sinusoidal modulation of the longitudinal coupling, the
monochromatic drive field with strength $\Omega_{\text{drv}}$ in
the original Hamiltonian, given by Eq.~(\ref{eq1}), induces two
coherent transition processes between the states
$|g,n+1\rangle\leftrightarrow|e,n\rangle$. The corresponding
transition rates are $C_{1\pm}$ (as shown with blue dashed and green solid
arrows). After compensating the dynamical Stark shift
 $\Delta_{\rm d}=\omega_q- \omega_{\rm pr}$, 
 the frequency separation between these two coherent
transitions is equal to the doubled coupling-modulation frequency,
i.e., $2\omega_{g}$. In the following discussions, we will show
that, assuming that the drive field is tuned properly, both
conventional and bichromatic EIT (or two-color EIT) can be
observed in such a hybrid system.

\section{Electromagnetically induced transparency}
In this section, we discuss the optical response to the probe
field via the standard master equation approach in our proposal.
The Born-Markov approximation is valid here.
Therefore, the system evolution is approximately described by the
Lindblad-type master equation
\begin{eqnarray}
\frac{d\rho (t)}{dt}&=&-i[H,\rho (t)]+\Gamma_{d} D[\sigma _{-}]\rho(t)+\Gamma_{\phi}D[\sigma _{z}]\rho(t) \notag \\
&&+(n_{\text{th}}+1)\kappa D[b]\rho(t)+n_{\text{th}}\kappa
D[b^{\dag}]\rho(t), \label{eq19}
\end{eqnarray}
where $D[A]\rho =(2A\rho A^{\dag }-A^{\dag}A\rho -\rho
A^{\dag}A)/2$ are the decoherence terms of the Lindblad
superoperator form, $\Gamma_{d}$ ($\Gamma_{\phi}$) is the decay
(pure dephasing) rate of the qubit, and $\kappa$ is the
relaxation rate of the mechanical mode due to its coupling with
a finite-temperature environment with a mean thermal
phonon number $n_{\text{th}}$.
Employing the methods of
Refs.~\cite{Astafiev2010,Abdumalikov10,Hoi11,Hoi12}, one can
approximately assume that the decay of the qubit is caused only by
the quantum noise in a 1D open line. The coherent drive and probe
fields for the qubit are also applied through the 1D transmission
line. Since the size of the qubit loop ($\thicksim\mu\rm{m}$) is
much smaller than the wavelength of the microwave drive
($\thicksim\rm{cm}$), we assume that the drive is
place-independent~\cite{Astafiev2010}.

The time-dependent atomic dipole moments can be expanded in terms
of the frequency Fourier components as~\cite{Yan13}
\begin{equation}
\langle\sigma _{-}(\omega^{\prime})\rangle
=\sum_{\omega^{\prime}=-\infty}^{\omega^{\prime}=\infty}
\langle\sigma _{-}(t)\rangle \exp(-i \omega^{\prime} t),
 \label{N1}
\end{equation}
where $\langle\sigma _{+}(t)\rangle$ can be found by numerically
solving the master equation (\ref{eq19}). Different from employing
a susceptibility to describe the optical response of an atomic
ensemble~\cite{Scully1997,Fleischhauer05}, here we should use the
reflection coefficient $r(\omega_{\text{pr}})$ to characterize the
electromagnetic response for the probe field of a single
atom~\cite{Wang17b}, which can be obtained via the following
relation~\cite{Astafiev2010,Abdumalikov10}
\begin{equation}
r(\omega_{\text{pr}})=-\frac{i\Gamma_{d}\langle\sigma
_{-}(\omega_{\text{pr}})\rangle}{2\Omega_{\text{pr}}}.
\label{eqS21}
\end{equation}
The real $\rm{Re}[r(\omega_{\text{pr}})]$ and imaginary
$\rm{Im}[r(\omega_{\text{pr}})]$ parts of the reflection
coefficient are related to the reflection and dispersion of a
single atom, respectively. In the following, we discuss how
$r(\omega_{\text{pr}})$ behaves under different drive and probe
conditions.

\subsection{Single-color EIT}
\begin{figure*}[t]
\centering
\includegraphics[width=1.00\linewidth]{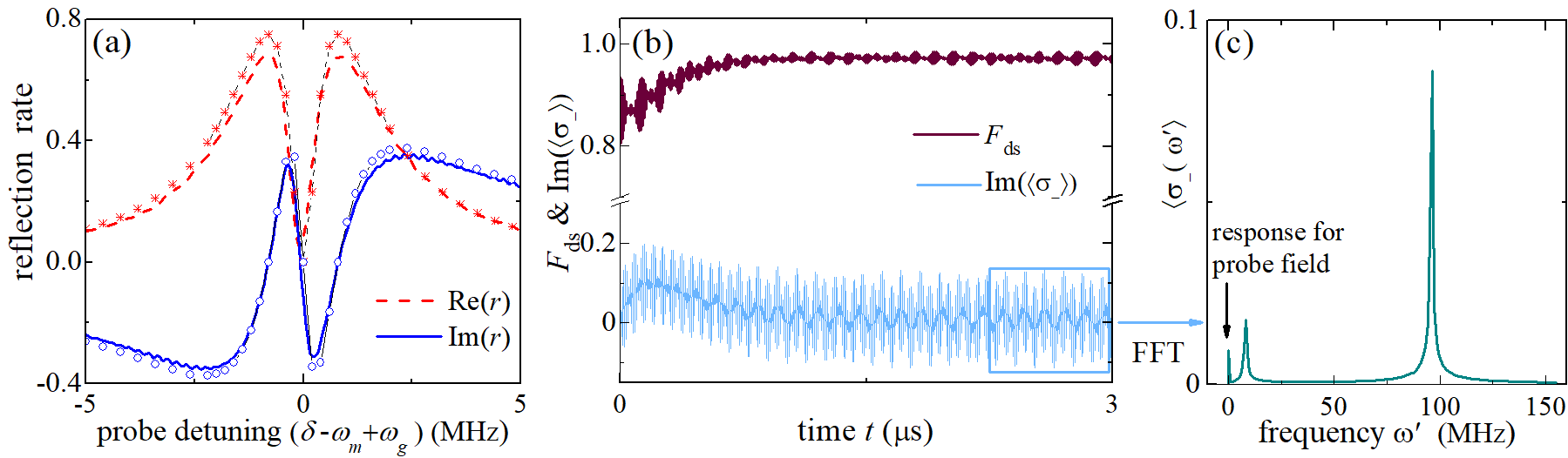}
\caption{Single-color EIT with $\delta_{s}=-\omega_{g}$. (a) The
real and imaginary parts of the reflection coefficient $r$,
$\text{Re}(r)$ (red dashed curve) and $\text{Im}(r)$ (blue solid
curve), as functions of the drive-probe detuning
$(\delta-\omega_{m}+\omega_{g})$ based on our numerical
simulation. The curves shown with stars and circles are plotted
according to analytical Eq.~(\ref{eqreff}). At
$\delta=\omega_{m}-\omega_{g}$, $\text{Re}(r)$ displays a dip with
$\text{Re}(r)\simeq0$. (b) Time evolutions of the dark-state
fidelity $F_{\rm ds}(t)$ (upper curves), given in Eq.~(\ref{eq19}), and
$\rm{Im}\langle \sigma_{-}\rangle$ (lower curves) of the qubit for the dip
position in (a). Note that $\text{Im}\langle \sigma_{-}\rangle$
oscillates around zero. (c) Employing the fast Fourier transform,
$\rm{Im}\langle \sigma_{-}\rangle$ in (b) is decomposed in the
frequency domain. Since our numerical calculations are performed
in the rotating frame of the probe field frequency, the dc (zero
frequency) component of $\rm{Im}\langle \sigma_{-}\rangle$
corresponds to the optical response of the probe field, which has
a low amplitude with $\rm{Im}\langle
\sigma_{-}(\omega'=0)\rangle\simeq9\times10^{-3}$, when EIT
occurs. The parameters are: $\omega_{m}/(2\pi)=100~\text{MHz}$,
$\Omega_{\text{drv}}/(2\pi)=10~\text{MHz}$, $\Omega_{\text{pr}}/(2\pi)=0.2~\text{MHz}$, $g_{0}/(2\pi)=8~\text{MHz}$,
$\omega_{g}/(2\pi)=4~\text{MHz}$, $\Gamma_{d}/(2\pi)=3~\text{MHz}$, $\Gamma_{\phi}/(2\pi)=0.25~\text{MHz}$,
$n_{\text{th}}=0$, and $\kappa/(2\pi)=1~\text{KHz}$.} \label{fig4}
\end{figure*}
The two split-sideband transitions are well-separated under the
condition $C_{\pm}\ll 2\omega_{g}$. Assuming that
$\delta_{s}=-\omega_{g}$ ($\delta_{s}=\omega_{g}$), only the
sideband transition $C_{1+}$ ($C_{1-}$) is on resonance, and we
reduce the Hamiltonian, in Eq.~(\ref{eq15}), by adopting the
rotating wave approximation, as follows
\begin{equation}
H_{\pm}=\big\{C_{1\pm}\sigma _{+}b -\Omega_{\text{pr}}\sigma _{+}
\exp[i(\omega_{m}\mp\omega_{g}-\delta)t]\big\}+\rm{H.c.}, \label{eq16}
\end{equation}
Under these conditions, the dynamical Stark shifts of the qubit
for the Hamiltonians $H_{\pm}$ are expressed as
\begin{eqnarray}
\Delta_{d\pm}&=&\sqrt{(\omega_{m}\mp\omega_{g})^{2}-4\Omega_{\text{drv}}^2}-(\omega_{m}\mp\omega_{g}) \notag \\
&&\simeq\frac{2\Omega_{\text{drv}}^{2}}{\omega_{m}\mp\omega_{g}}\simeq\frac{2\Omega_{\text{drv}}^{2}}{\omega_{m}}.
\label{eq17}
\end{eqnarray}

To observe single-color EIT, we assume that the sideband
transition $C_{1+}$ is resonantly selected. Moreover, to suppress
the transition $C_{1-}$, the condition $C_{\pm}\ll 2\omega_{g}$ should always be
satisfied to ensure the validity of the rotating wave approximation.
When the probe-drive
detuning satisfies the condition $\delta=\omega_{m}-\omega_{g}$,
as shown in Fig.~\ref{fig3}, the transitions are represented by
the red and blue arrows. As a result, the
evolution of the system is approximately described by the
time-independent Hamiltonian
\begin{equation}
H_{+}=\big(C_{1+}\sigma _{+}b-\Omega_{\text{pr}}\sigma
_{+}\big)+\rm{H.c.} \label{eq18}
\end{equation}
Assuming that the NAMR has a high-quality factor and the condition
$\min\{C_{1+},\omega_{\text{pr}},\Gamma_{d}\}\gg n_{\text{th}}\kappa$ is
satisfied, the decoherence process of the NAMR can be neglected.
The effective Hamiltonian $H=H_{+}$, together with the rapid decay
of the qubit, drives the system into the following dark
state~\cite{Sun06,Wang16e}:
\begin{equation}
|\Psi_{\rm ds}\rangle=e^{-|\lambda^{2}|/2}\sum_{n}
\frac{\lambda^{n}}{\sqrt{n!}}|g,n\rangle=|g\rangle|\alpha_{\lambda}\rangle,
\label{DarkState}
\end{equation}
where $\lambda=\Omega_{\text{pr}}/C_{1+}$ and
$|\alpha_{\lambda}\rangle$ is a coherent state.

In a typical EIT system, the probe field is weak compared with the
control field, i.e., $\Omega_{\text{pr}}\ll C_{1+}$. Therefore
$\lambda\ll1$, and we can use the states $|g,0\rangle$,
$|g,1\rangle$, and $|e,0\rangle$ to describe the transitions
governed by Eq.~(\ref{eq16}). The relation between these three
states is similar to a $\Lambda$-type EIT system. The reflection
coefficient of the probe field for $H=H_{+}$ in Eq~(\ref{eq16}) is
expressed as~\cite{Scully1997,Fleischhauer05}:
\begin{equation}
r_{\rm{eff}}(\omega_{\text{pr}})=\frac{\Gamma_{d}}{2\Gamma_{f}-2i(\delta-
        \omega_{m}+\omega_{g})+\frac{4C_{1+}^{2}}{\kappa-2i(\delta-\omega_{m}+\omega_{g})}},
\label{eqreff}
\end{equation}
where $\Gamma_{f}=\Gamma_{d}/2+2\Gamma_{\phi}$ is the total
dephasing rate. Equation~(\ref{eqreff}) indicates that the
mechanical decay rate $\kappa$ determines the width of the EIT
window. Thus, very narrow EIT windows can be observed in our
proposal by adopting a high quality-factor NAMR with
$\Gamma_{f}\gg \kappa$.
If the NAMR is implemented by a
carbon nanotube, the quality factor can be extremely
high~\cite{Moser14}. Thus, we assume that the NAMR is vibrating at
mechanical frequency $\omega_{m}/(2\pi)=100~\text{MHz}$ with
$\kappa/(2\pi)=1~\text{KHz}$ (see~\cite{Laird11,Benyamini14}).
The Rabi frequencies of the two coherent drives are
$\Omega_{\text{drv}}/(2\pi)=10~\text{MHz}$ and $\Omega_{\text{pr}}/(2\pi)=0.2~\text{MHz}$, respectively. For the
modulated coupling $g(t)$, we set $g_{0}/(2\pi)=8~\text{MHz}$ and
$\omega_{g}/(2\pi)=4~\text{MHz}$, as discussed in Sec.~II B. According to Eq.~(\ref{eq13}), the effective
sideband transition rate is $C_{1+}/(2\pi)=0.8~\text{MHz}$.
For a superconducting qubit interacting with an open one-dimensional transmission line,
the energy relaxation and dephasing rates are about MHz~\cite{Astafiev2010,Anisimov2011}, and here we set $\Gamma_{d}/(2\pi)=3~\text{MHz}$
and $\Gamma_{\phi}/(2\pi)=0.2~\text{MHz}$, respectively.
Employing these
parameters, in Fig.~\ref{fig4}(a) we plot
$\rm{Re}[r(\omega_{\text{pr}})]$ (red solid curve) and
$\rm{Im}[r(\omega_{\text{pr}})]$ (blue solid curve) changing with
the detuning $(\delta-\omega_{m}+\omega_{g})$ by numerically
solving the master equation with the original Hamiltonian in
Eq.~(\ref{eq2}) (rotating at the probe frequency). Moreover, the
analytical form for $r_{\rm{eff}}$, given in Eq.~(\ref{eqreff}),
is also plotted with the curves shown with symbols.

In Fig.~\ref{fig4}(a), one can see a single EIT dip with
$\rm{Re}[r(\omega_{\text{pr}})]\simeq0$ around $\delta=\omega_{m}-\omega_{g}$.
Different from conventional atomic EIT, the control field here is
not a semiclassical coherent drive, but a parametrically modulated
coupling inducing a sideband transition $C_{1+}$. Moreover, we
find that our analytical and numerically results of the optical
response match well with each other, indicating that the
Hamiltonian Eq.~(\ref{eq18}) can effectively describe the
transition relation of the single-window EIT in
Fig.~\ref{fig4}(a).

Defining the fidelity
\begin{equation}
F_{\rm ds}(t)=\langle\Psi _{\rm ds}|\rho(t)|\Psi _{\rm ds}\rangle
\end{equation}
for the dark state in Eq.~(\ref{DarkState}), Fig.~\ref{fig4}(b)
depicts the time evolution of this fidelity and
$\rm{Im}\langle\sigma _{-}(t)\rangle$ with
$\delta=\omega_{m}-\omega_{g}$ [the dip position in
Fig.~\ref{fig4}(a)]. The numerical results clearly show that the
system is rapidly steered into its dark state with the steady
fidelity $F_{\rm ds}\simeq98.5\%$. Note that $\rm{Im}\langle\sigma
_{-}(t)\rangle$ oscillates in time, and it contains many frequency
components. The fast Fourier transform of $\rm{Im}\langle\sigma
_{-}(t)\rangle$ is shown in Fig.~\ref{fig4}(c), which exhibits
three main peaks in the low-frequency regime. The first peak at
$\omega^{\prime}=0$ (i.e., the dc component) corresponds to the
optical response of the probe field. This peak has very low
amplitude due to EIT.
\begin{figure}[t]
        \centering \includegraphics[width=8.6cm]{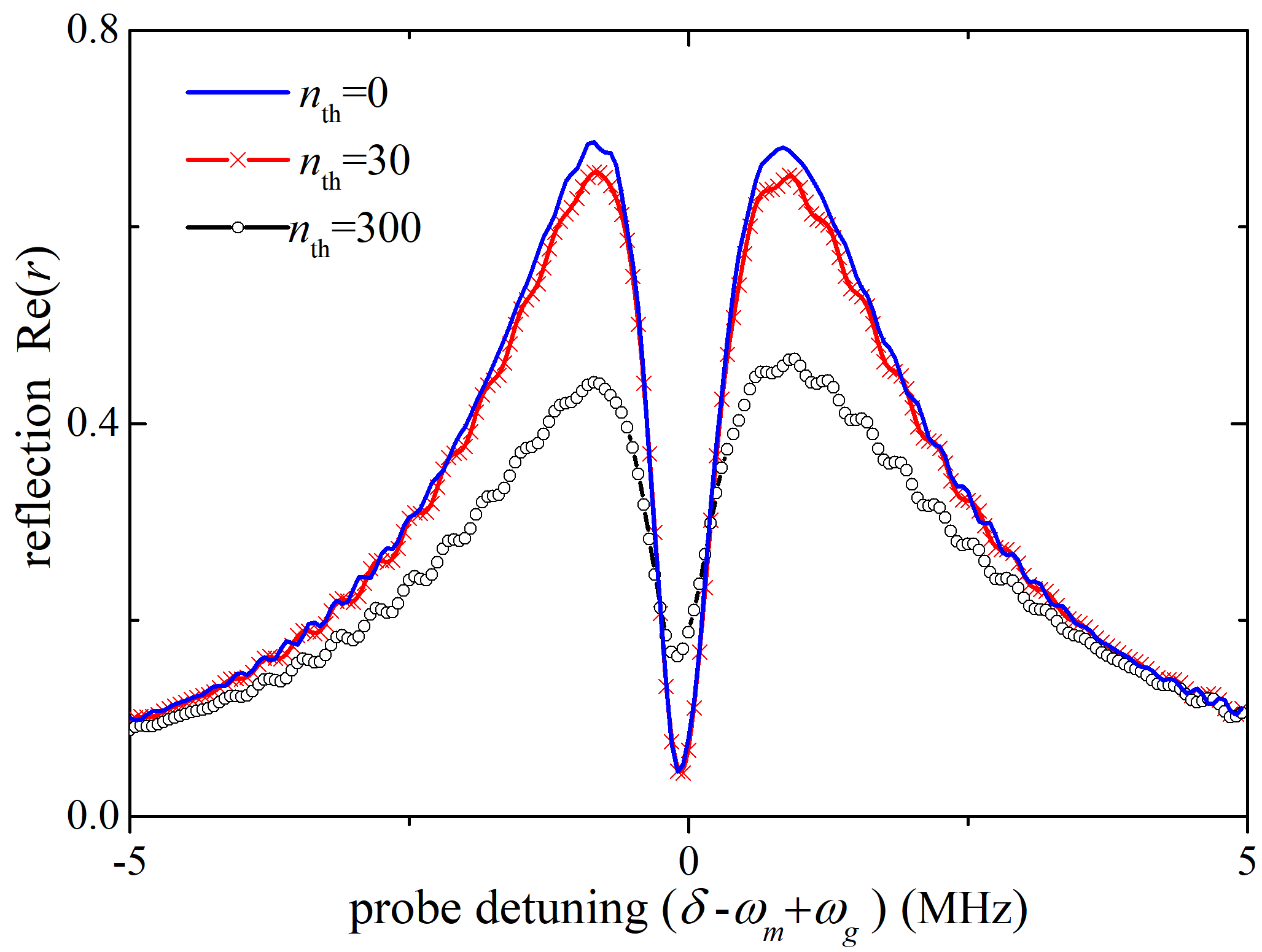} \caption{
                The real part of the reflection coefficient $r$,
                $\text{Re}(r),$ as a function of the probe detuning
                $(\delta-\omega_{m}+\omega_{g})$ for $n_{\text{th}}=0$ (blue solid curve),
                 $n_{\text{th}}=30$ (red cross) and $n_{\text{th}}=300$ (black circle).
                 Other parameters employed here are
                the same as those in Fig.~\ref{fig4}.}
        \label{fig5}
\end{figure}

Note that the NAMR has much lower eigenfrequency than  that of the qubit, so
it might couple with a finite-temperature environment with
thermal phonon number $n_{\text{th}}$.
Once the effective decay rate  $n_{\text{th}}\kappa$
of the NAMR is close to $\min\{C_{1+},\omega_{\text{pr}},\Gamma_{d}\}$, the dark state, given in
Eq.~(\ref{DarkState}),  is destroyed by  thermal noise.
In Fig.~\ref{fig5}, we show how the reflection rate behaves around the EIT window for different values of $n_{\text{th}}$.
We find that, when increasing
$n_{\text{th}}$, the EIT dip becomes shallower with a wider EIT window, indicating that  EIT
is influenced by thermal phonons.
However, only when $n_{\text{th}}$ is $\ge300$, this
damage effect is clearly apparent.
For the case with $n_{\text{th}}=30$, the EIT  effect differs only slightly from the zero-temperature case ($n_{\text{th}}=0$).
Given that the proposed system is placed in a
dilution refrigerator at $20~\text{mK}$,  the corresponding
thermal phonon occupation is $n_{\text{th}}\simeq4$, and the EIT
condition $\min\{C_{1+},\omega_{\text{pr}},\Gamma_{d}\}\gg n_{\text{th}}\kappa$ can be easily satisfied.
Therefore, the thermal noise affecting the NAMR can be neglected in our discussions.

\begin{figure}[t]
        \centering \includegraphics[width=8.4cm]{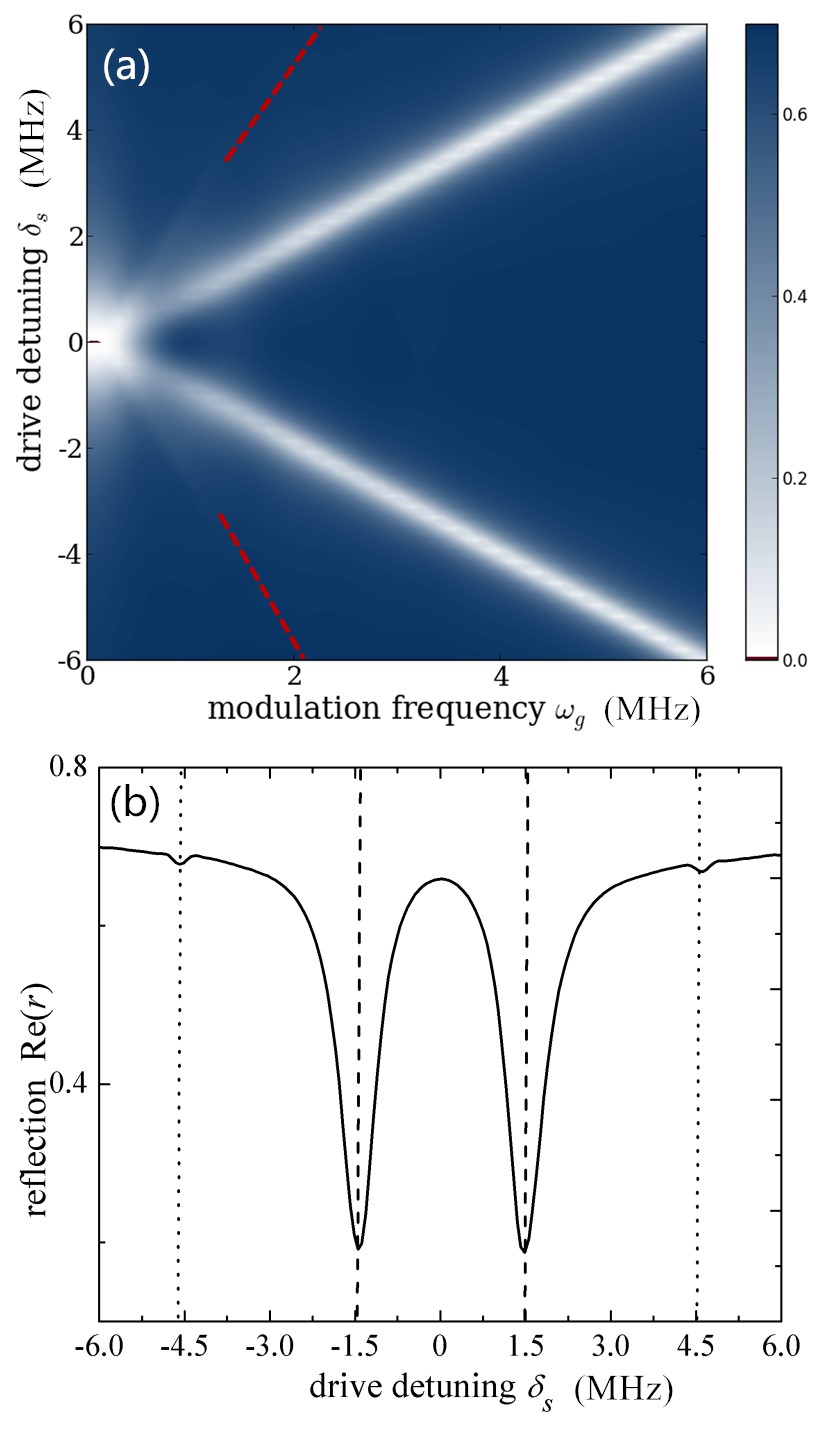} \caption{(a)
                The reflection rate $\text{Re}(r)$ of the resonant probe field
                versus the drive detuning $\delta_{s}$ and the coupling-modulation
                frequency $\omega_{g}$. At $\omega_{g}=0$, there is only one
                sideband drive frequency for the single-color EIT window. With
                increasing $\omega_{g}$, the single dip for the EIT drive field is
                split into two apparent dips at $\delta_{s}\simeq \pm\omega_{g}$,
                and two less apparent dips at $\delta_{s}\simeq \pm 3\omega_{g}$
                as additionally indicated by the dotted red lines. (b) The
                cross-section of figure (a) at $\omega_{g}/(2\pi)=1.5~\text{MHz}$. Two
                shallow dips induced by the third-order terms $C_{3\pm}$ can be
                observed at $\delta_{s}/(2\pi)=\pm 4.5~\text{MHz}$. Other parameters employed here are
                the same as those in Fig.~\ref{fig4}.} \label{fig6}
\end{figure}

The single-color EIT demonstrated here is quite different from the
case when the qubit-NAMR longitudinal coupling is
constant~\cite{Lijj12,Wang16e}, where the drive-probe detuning
should be fixed and exactly equal to the eigenfrequency of the
NAMR. In the case studied here with a modulation interaction, the
drive-probe detuning is continuously changed by varying the
coupling-modulation frequency $\omega_{g}$, which is analogous to
changing the frequency difference between two metastable states in
the EIT system. This can be clearly seen in Fig.~\ref{fig6}(a),
where we fix the probe field to be resonantly applied to the
qubit, and plot the optical response
$\text{Re}[r(\omega_{\text{pr}})]$ by changing the drive detuning
$\delta_{s}$ and the coupling-modulation frequency $\omega_{g}$.
At $\omega_{g}\simeq0$, there is only one drive-probe detuning
position for the single EIT window
($\tilde{\Delta}\simeq\omega_{m}$). Intriguingly, when we start to
increase the frequency $\omega_{g}$ from zero, the single dip of
the reflection rate as a function of the drive detuning splits
into two apparent dips separated by $~2\omega_{g}$, which
corresponds to the transition rates $C_{1+}$ and $C_{1-}$, respectively. Two possible
drive frequencies can induce single-color EIT. Thus, modulating
the qubit-NAMR coupling at frequency $\omega_{g}$ is equivalent to
replacing the single NAMR with two frequency tunable NAMRs with
frequencies $\omega^{\prime}_{m}=\omega_{m}\pm\omega_{g}$.

Moreover, we find \emph{two shallow dips} along the dashed lines
in Fig.~\ref{fig6}(a), which can be seen clearly in the
cross-section plot Fig.~\ref{fig6}(b) by fixing the
coupling-modulation frequency at $\omega_{g}/(2\pi)=1.5~\text{MHz}$. The relation
between drive detuning and the modulation frequency is
approximately given by $\delta_{s}\simeq \pm3\omega_{g}$. These
two transparent dips result from the third-order resonant
couplings $C_{3\pm}$ in Eq.~(\ref{eq14}). However, due to
extremely low rates, these two dips are much shallower than those
induced by $C_{1\pm}$.
\subsection{Two-color EIT}
In previous discussions, we found that only one sideband
transition was dominant by setting  the coupling-modulation
frequency $\delta_{s}=\pm\omega_{g}$. However, by assuming that
$\delta_{s}=0$, the two sideband transitions (with strengths
$C_{\pm}$) correspond to the same detuning, so both of these
should be considered equally. The Hamiltonian in Eq.~(\ref{eq15})
is now reduced to
\begin{equation}
H=\sum_{j=\pm}\big[C_{1j}\sigma _{+}b e^{ j i\omega_{g}t}
-\Omega_{\text{pr}}\sigma _{+} e^{i(\omega_{m}-\delta)
t}\big]+\rm{H.c.} \label{eq20}
\end{equation}
As shown in Fig.~\ref{fig3}, both sideband transitions $C_{1+}$
and $C_{1-}$ are detuned by $\omega_{g}$. The energy-level
transition relation is similar to two-color EIT with two control
fields studied in Refs.~\cite{Wang06,Moiseev06,Yan13}. Two
transparent windows for the probe field were observed. In contrast
to two-color EIT studies in atomic
systems~\cite{Wang03,Moiseev06,Liuy2012,Yan13,Liuy2012}, only one
(not two) coherent drive is employed here. The two split
transparent windows result from a monochromatic modulation of the
longitudinal coupling. Indeed, if $g(t)$ contains $N$
well-separated frequency components, $2N$ windows of EIT can be
observed.
\begin{figure}[t]
\centering \includegraphics[width=8.8cm]{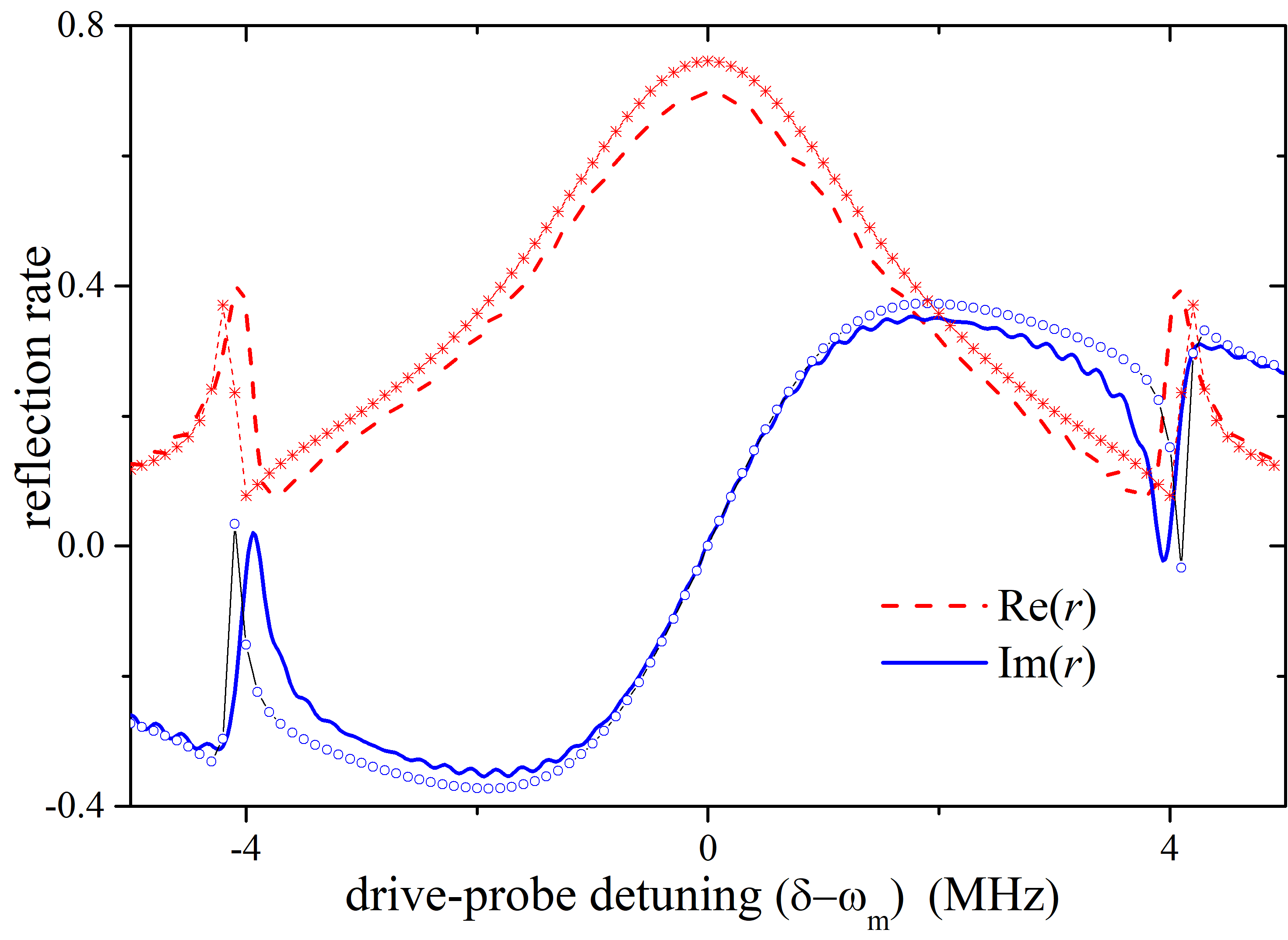}
\caption{Two-color EIT under the condition $\delta_{s}=0$. The
real and imaginary parts of the reflection coefficient $r$,
$\text{Re}(r)$ (red dashed curve) and $\text{Im}(r)$ (blue solid
curve), versus the drive-probe detuning $(\delta-\omega_{m})$
based on our numerical simulations. The analytical average optical
response $r_{c}$, as given in Eq.~(\ref{eqr2}), is plotted with
symbols. Two transparent dips can be found at $\pm\omega_{g}$.
Other parameters adopted here are the same as those in
Fig.~\ref{fig4}.} \label{fig7}
\end{figure}

Analogous to a conventional $\Lambda$-type EIT system, if we
consider only one detuning sideband transition (either $C_{1-}$ or
$C_{1+}$) of detuning $\omega_{g}$ in Eq.~(\ref{eq20}), the
reflection coefficient becomes
\begin{equation}
r_{\rm{eff}\pm}(\omega_{\text{pr}})=\frac{\Gamma_{d}}{2\Gamma_{f}
-2i(\delta-\omega_{m})+\frac{4C_{1\pm}^{2}}{\kappa-2i(\delta-\omega_{m}\mp\omega_{g})}},
\label{eqreff2}
\end{equation}
with a real part, which has dip positions at
$\delta-\omega_{m}=\pm\omega_{g}$. If both sideband transitions
$C_{1-}$ and $C_{1+}$ occur with a symmetric detuning, the optical
response for the probe field should combine these two EIT effects.

In Fig.~\ref{fig7}, the imaginary and real parts of the reflection
coefficient $r(\omega_{\text{pr}})$ are plotted based on our
numerical simulations (the blue solid and red dashed curves). Around
$\delta-\omega_{m}\simeq\pm\omega_{g}$, the two EIT windows emerge
with typical anomalous dispersion curves of negative slope.
Therefore, by \emph{applying a single-drive field, we can simultaneously control the
transparency for two microwave fields} when their frequency
separation equals $2\omega_{g}$. Moreover, following
Eq.~(\ref{eqreff2}), the analytical mean optical response
$r_{c}=-(i\Gamma_{d}\langle\sigma
_{-}\omega_{\text{pr}})\rangle/(2\Omega_{\text{pr}})$ can be approximately expressed as
\begin{equation}
r_{c}=\frac{r_{\rm{eff}+}+r_{\rm{eff}-}}{2}, \label{eqr2}
\end{equation}
which is plotted by the curves with symbols (either stars or
circles) in Fig.~\ref{fig7}. Interestingly, we find that the
analytical results in Eq.~(\ref{eqr2}) can approximately describe
the \emph{joint two-color EIT}. The optical response can be viewed
as a combined effect of two isolated EIT effects with the same
drive detuning, and their interference is negligible, given that
$C_{1\pm}$ are much weaker compared with $2\omega_{g}$. As shown
in Eq.~(\ref{eqC1pm}), the sideband transition $C_{1+}$ is greater
than $C_{1-}$, and the rate difference becomes more apparent when
increasing the modulation frequency $\omega_{g}$. When
$\omega_{g}$ is large enough, the two dips in Fig.~\ref{fig7} are
not symmetric anymore.
\begin{figure}[t]
\centering \includegraphics[width=8.5cm]{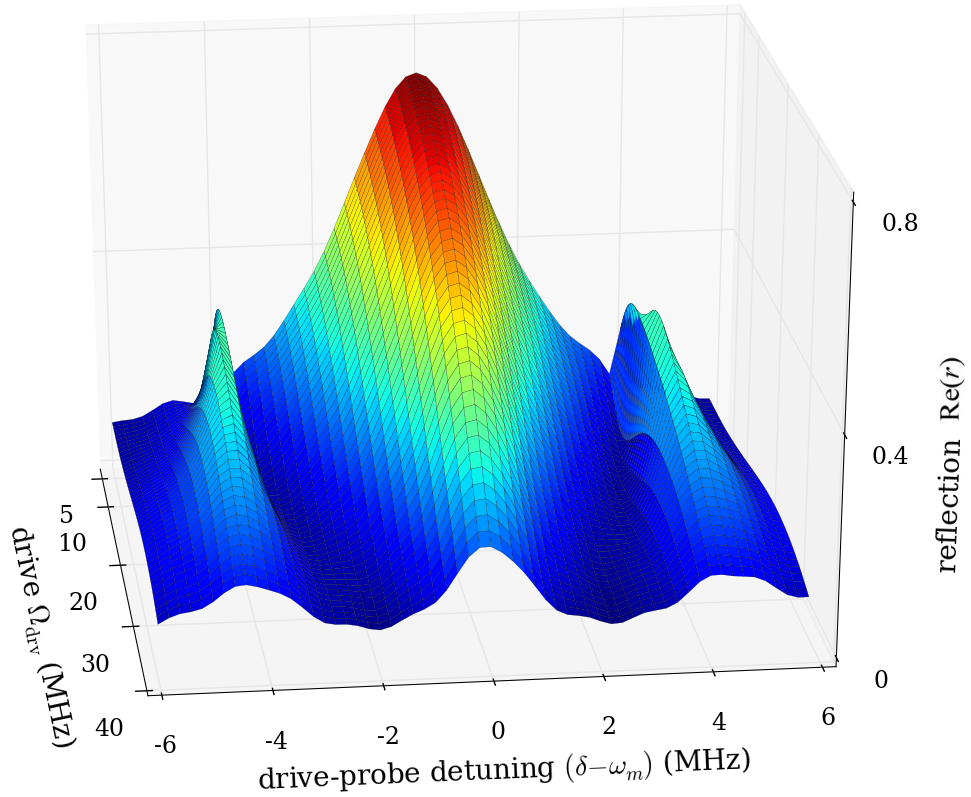} \caption{The
real part of the reflection coefficient, $\text{Re}(r)$, versus
the drive-probe detuning $(\delta-\omega_{m})$ and drive strength
$\Omega_{\rm{drv}}$ for the two-color EIT. When increasing
$\Omega_{\rm{drv}}$, the two EIT dips located at $\pm\omega_{g}$,
become wider and closer to each other, indicating that the
two-color EIT is gradually destroyed by the interference of
the two sideband transitions. Other parameters adopted here are
the same as those in Fig.~\ref{fig4}. } \label{fig8}
\end{figure}

According to Eq.~(\ref{eq13}), both transition rates $C_{1-}$ and $C_{1+}$ linearly
increase with increasing $\Omega_{\text{drv}}$. In
Fig.~\ref{fig8}, we plot the reflection rate
$\text{Re}[r(\omega_{\text{pr}})]$ as a function of the drive
strength $\Omega_{\text{drv}}$ and probe detuning
$(\delta-\omega_{m})$. We find that, with increasing
$\Omega_{\text{drv}}$, the two EIT dips become wider and closer
due to strong sideband transition rates $C_{1\pm}$. The two
isolated transparent windows affect each other, and they tend to
merge. When $C_{1\pm}$ is comparable with the frequency separation
$2\omega_{g}$, the relation $C_{\pm}\ll \omega_{g}$ is not valid
any more, and this two-color EIT almost disappears. If the
two sideband transitions are not well-separated by frequency
detuning, Eq.~(\ref{eqr2}) cannot effectively describe the optical
response, and the two EIT windows are destroyed.

\section{Discussion and conclusions}

In this work, we considered a hybrid system consisting of a SQUID
embedded with a NAMR. We first showed an example of how to achieve
an unconventional parametrically-modulated longitudinal
interaction between a flux (transmon) qubit and the NAMR. Then, we
derived an effective Hamiltonian, which leads to a first-order
sideband transition, and found that the coupling modulation
significantly changes the dynamics of the hybrid system. A single
sideband drive is split under a sinusoidal modulation of the
coupling terms. Indeed, the frequency components of the modulation
directly determine this splitting. If the modulation is more
complex, then more interesting phenomena can be observed.

By applying a resonant probe field, we found that both single- and
two-color EIT can be observed. The modulation of the interaction
provides another control method for these EIT effects. For the
single-color EIT, the drive-probe detuning is not necessarily
equal to the NAMR frequency, but can be conveniently tuned by
changing the modulation frequency. For the two-color EIT, the
double transparent windows occur due to the splitting of sideband
transitions, and their distance is determined by the modulation
frequency. Compared with the usual predictions of two-color EIT in
atomic systems, here there is only one drive (control) field.
Moreover, it is possible to modify and extend our results to study
EIT and Autler-Townes splitting~\cite{Anisimov2011,Peng2014,Sun2014}.

As discussed in Ref.~\cite{Wang17b}, for systems with longitudinal
interaction, an EIT induced by second-order sideband transitions
can also be observed. By considering the modulation of such
longitudinal interaction, one might observe multi-color EIT
induced by higher-order sideband transitions.

We hope that our results could not only be helpful for studying
the dynamics for a system with time-dependent longitudinal
coupling, but also can find applications in microwave
photonics~\cite{Gu2017,Kockum2018} (including vacuum-induced nonlinear
optics~\cite{Kockum2017,Stassi2017}) and quantum information
processing~\cite{Wendin2017} with SQCs.

\section*{Acknowledgements}

X.W. and H.R.L. were supported by the Natural Science Foundation
of China under Grant No. 11774284. A.M. and F.N. acknowledge the
support of a grant from the John Templeton Foundation. F.N. is
partially supported by the MURI Center for Dynamic
Magneto-Optics via the AFOSR Award No.~FA9550-14-1-0040, the Army
Research Office (ARO) under grant number 73315PH, the AOARD grant
No.~FA2386-18-1-4045, the CREST Grant No.~JPMJCR1676, the IMPACT
program of JST, the RIKEN-AIST Challenge Research Fund, and the
JSPS-RFBR grant No.~17-52-50023.

\end{document}